\documentclass[]{iopart}
\usepackage{iopams}
\usepackage{amsthm}
\usepackage{setstack}
\usepackage{graphicx}
\usepackage{epsfig}
\usepackage{tensor}
\usepackage[usenames]{color}
\usepackage{cite}
\usepackage{pifont}

\theoremstyle{plain}
\theoremstyle{definition}

\usepackage[colorlinks=true,linkcolor=blue,citecolor=blue,urlcolor=blue]{hyperref}
\begin{document}

\title[Newton-Cartan Gravity in Noninertial Reference Frames]{Newton-Cartan Gravity in Noninertial Reference Frames}

\author{Leo Rodriguez$^1$, James St.\ Germaine-Fuller$^2$, Sujeev Wickramasekara$^2$}

\address{$^1$ Physics Department, Assumption College, Worcester, MA 01609}
\address{$^2$ Physics Department, Grinnell College, Grinnell, IA 50112}
\eads{\mailto{ll.rodriguez@assumption.edu}, \mailto{stgermai@grinnell.edu}, \mailto{wickrama@grinnell.edu}}
\begin{abstract}
We study properties of Newton-Cartan gravity under transformations into all noninertial, nonrelativistic reference frames. The set of these transformations has the structure of an infinite dimensional Lie group, called the Galilean line group, which contains as a subgroup the Galilei group. We show that the fictitious forces of noninertial reference frames are naturally encoded in the Cartan connection transformed under the Galilean line group. These noninertial forces, which are coordinate effects, do not contribute to the Ricci tensor which describes the curvature of  Newtonian spacetime. We show that only the $00$-component of the Ricci tensor is non-zero and equal to ($4\pi$ times) the matter density in any inertial or noninetial reference frame and that it leads to what may be called Newtonian ADM mass. While the Ricci field equation and Gauss law are both fulfilled by the same physical matter density in inertial and linearly accelerating reference frames, there appears a discrepancy between the two in rotating reference frames in that Gauss law holds for an effective mass density that differs from the physical matter density. This effective density has its origin in the simulated magnetic field that appears in rotating frames, highlighting a rather striking difference between linearly and rotationally accelerating reference frames.  We further show that the dynamical equations that govern the simulated gravitational and magnetic fields  have the same form as  Maxwell's equations, a surprising conclusion given that these equations are well-known to obey special relativity (and $U(1)$-gauge symmetry), rather than Galilean symmetry. 
\end{abstract}

\maketitle
\section{Introduction}\label{sec1} 
The purpose of this article is to study the covariance of Cartan's geometric formulation of Newtonian gravity under 
transformations into reference frames moving with arbitrary accelerations, 
both linear and rotational. These transformations have the structure of an infinite dimensional Lie group that has been called the Galilean line group,
\cite{2012AnPhy.327.2310M} $\mathbb{G}$. We show that the essential geometric content of Newton-Cartan gravity remains intact under this group.
In particular, only the $00$-component of the Ricci tensor is non-zero in all reference frames and equal to ($4\pi$ times) the matter density. 
As such, the Ricci field equation $R_{00}=4\pi\rho$, which describes the curvature of the Newtonian spacetime,  is a scalar under $\mathbb{G}$.  
The equation of motion of a test particle, which can be recast as the autoparallel curve defined by Cartan's connection, 
also has well-defined transformation properties  and fictitious forces are naturally generated from the transformations of this connection under $\mathbb{G}$.
 A particularly interesting feature of this geometric formulation of noninertial effects is that the 
connection components that encode the Coriolis effect can be used to define a simulated magnetic vector potential, a property that may be anticipated 
from the structural similarity between the Coriolis force $2m\boldsymbol{v}\times\boldsymbol{\omega}$ and the Lorentz force $e\boldsymbol{v}\times\boldsymbol{B}$. 
Likewise, the connection components that include the centrifugal and Euler terms, which depend on the position rather than velocity, have a natural reading as a simulated 
gravitational field. We show that together, the fictitious forces have the structure of a simulated gravitomagnetic field which, much like the electromagnetic field, is described by Maxwell's equations, 
while the equation of motion of a nonrelativistic test particle in this gravitomagnetic field has the same form as the equation of motion of a particle moving 
under the Lorentz force (with mass as the coupling constant). Not surprisingly, the field equations for our gravitomagnetic field has $U(1)$-gauge symmetry, an automatic consequence 
of introducing vector and scalar potentials. Being of the same form as Maxwell's equations, these field equations 
can also be considered a tensor equation under Lorentz transformations. 
In this regard,  our study has an interesting parallel to Dyson's \cite{Dyson} account of Feynman's proof that Maxwell's equations can be derived 
from Newton's law of motion and Heisenberg's canonical commutation relations. Since these commutation relations are in fact a consequence of 
unitary projective representations of the Galilei group, the crux of the Dyson construction is rather similar to ours: start with a Galilean theory and arrive 
at a special relativistic theory. However, we note that a physical interpretation of the Lorentz symmetry of simulated gravitomagnetic fields is not straightforward and perhaps 
not tenable owing to the noninertial character of the reference frames involved. 

Further, our study highlights interesting differences between linearly and rotationally accelerating reference frames. The simulated magnetic field and its associated vector potential 
come into being only in rotating reference frames, never in linearly accelerating frames. Consequently, the equation of motion remains
form invariant and the Ricci field equation remains consistent with Gauss' law in linearly accelerating reference frames while these properties do not hold in rotating 
reference frames. Thus, this study provides a fresh perspective on an old perception: rotationally accelerating reference frames are different in character from linearly accelerating  ones and, 
from a gravitational point of view, the latter are not all that different from inertial reference frames.

The motivation of this study derives from some recent work on formulating quantum mechanics in noninertial reference frames on the basis of unitary cocycle representations of $\mathbb{G}$\cite{PhysRevLett.111.160404,2012AnPhy.327.2310M,2014AnPhy.340...94K,2013AnPhy.336..261K,2013arXiv1305.5021K}.
These studies have provided several new insights into Galilean quantum mechanics.
In particular, they have shown that the Wigner-Bargmann notion that an elementary particle is defined by a unitary irreducible representation of the relevant spacetime symmetry group may be extended to noninertial reference frames. A rather remarkable property of this extension of the Wigner-Bargmann program is that, with the exception of a few, the representations of $\mathbb{G}$ that may be used to define a particle lead to violations of the equivalence principle at the quantum level, a consequence of the group cohomology of $\mathbb{G}$ which is much richer than that of the Galilei or Poincar\'e groups. 
Further, the representations of $\mathbb{G}$ provide the means to rigorously derive the above mentioned simulated magnetic fields \cite{PhysRevLett.111.160404} in the quantum case. However, the most interesting mathematical feature of quantum mechanics in noninertial frames is that it requires a certain non-associative \emph{loop} extension of $\mathbb{G}$ when rotating reference frames are present, while linear accelerations can be well accommodated within a more traditional group extension. This discrepancy between rotational and linear accelerations in Galilean quantum mechanics parallels that in the classical Newton-Cartan theory studied here. 

The organization of the paper is as follows. In Section~\ref{sec2}, we briefly review  Cartan's geometric formulation of Newton's gravity. In Section~\ref{sec:NCGroup}, we introduce the Galilei and Galilean line groups and study the transformation properties of the Newton-Cartan theory under these groups.  The analysis of the Newtonian limit of ADM mass, done in the context of a Reissner-Nordstr\"om black hole, is the subject of Section~\ref{sec4}. We offer some concluding remarks in Section~\ref{sec5} and present a few calculational details in \ref{app:1} and~\ref{app:2}. 
\section{Newton-Cartan gravity}\label{sec2}
Traditionally, Newton's Gravity is thought of in terms of its inverse square force law, 
\begin{eqnarray}
\label{eq:fng}
F_{grav}=\frac{GMm}{r^2},
\end{eqnarray}
between two point masses $M$ and $m$, where $G$ is Newton's gravitational constant.  However, just like Coulomb's force law in electrodynamics, 
\eref{eq:fng} can be recast as Gauss' law:
\begin{eqnarray}
\label{eq:glng}
\nabla\cdot \vec g=-4\pi \rho,
\end{eqnarray}
where $\vec g$ is the gravitational field strength and the minus sign reflects the fact that test particles are pulled toward the center of the mass distribution $\rho$. 
In \eref{eq:glng}, we have used natural units $G=c=1$, a choice we will make for the remainder of the paper. The totality of Newton's gravity, for time independent fields, 
may be summed up by the additional path independence property of $\int_{\vec{r}_1}^{\vec{r}_2} \vec g\cdot d\vec{r}$:
\begin{eqnarray}
\label{eq:ng}
\cases{
\nabla\cdot \vec g=-4\pi \rho&Gauss-Law\\
\nabla\times \vec g=0&$\vec g$ is conservative
}
\end{eqnarray}
Just as in electrostatics, path independence allows us to introduce the Newtonian gravitational field (potential) $\Phi$ such that $ \vec g=-\nabla\Phi,$ reducing the two equations in \eref{eq:ng} to a single Poisson equation, 
\begin{eqnarray}
\label{eq:ngpeq}
\nabla^2\Phi=4\pi \rho,
\end{eqnarray}
which completely describes the gravitational field arising from matter density $\rho$. The motion of a test particle of mass $m$ in this gravitational field is determined 
by Newton's second law, 
\begin{eqnarray}
\label{eq:eqmng}
F=m\frac{d^2\vec{x}}{dt^2}=-m\nabla\Phi.
\end{eqnarray}
Implied in \eref{eq:eqmng} is the \emph{equivalence principle}, namely that the test particle couples to the gravitational field by means of its inertial mass. 
Consequently, mass drops out of \eref{eq:eqmng}, the acceleration of any test particle becomes the same as the field $\vec{g}$, and 
all particles have the same set of trajectories. Though an incomplete formulation of gravity at all energy scales, Newton's theory still encodes a wealth of information about the nature of spacetime.

After Einstein worked out his general theory of relativity (GR), Cartan showed that Newtonian gravity, too, can be cast in the form of a geometric theory. Just as in GR, the key is again the equivalence 
principle which allows any solution to the equation of motion \eref{eq:eqmng} to be recast as an autoparallel curve. 
To see this,  let us introduce a universal (Galilean-affine) time $\tau = \lambda t+b$, where $\lambda$ and $b$ are constants,  and rewrite \eref{eq:eqmng} as
\begin{eqnarray}
\label{Newton}
\frac{d^2 x^i}{d\tau^2} + \delta^{ij} \frac{\partial \Phi}{\partial x^j} \left(\frac{dt}{d\tau}\right)^2 = 0.
\end{eqnarray}
Here, we have used the  Euclidean metric $\delta^{ij}$
to raise (and lower) spatial indices and adopted Einstein summation notation, 
conventions we will follow in the rest of the paper. Our Latin indices run over 1, 2, and 3 
while Greek indices run over 0, 1, 2, and 3.

A comparison of \eref{Newton} with the general autoparallel equation 
\begin{eqnarray}
\label{eq:gdeq}
\frac{d^2 x^\mu}{d \tau^2}+\Gamma^\mu_{\alpha\beta} \frac{dx^\alpha}{d\tau}\frac{dx^\beta}{d\tau} = 0
\end{eqnarray} 
shows that \eref{Newton}  describes an autoparallel curve where the only non-zero connection components are
\begin{eqnarray}
\label{eq:ncc}
\Gamma^i_{00} = \delta^{ij} \frac{\partial \Phi}{\partial x^j} = -g^i.
\end{eqnarray}

The non-vanishing affine connection and its relation to the gravitational potential implies that, just as in GR, 
matter introduces a curvature to the Newtonian spacetime and that the trajectories of test particles under the influence of gravity are simply autoparallel curves in this curved spacetime.    
In fact, the Ricci tensor, 
\begin{eqnarray}\label{eq:ricci}
R_{\alpha \beta} = \partial_\rho \Gamma^\rho_{\beta\alpha} -  \partial_\beta \Gamma^\rho_{\rho\alpha} + \Gamma^\rho_{\rho\lambda}\Gamma^\lambda_{\beta\alpha} - 
\Gamma^\rho_{\beta\lambda}\Gamma^\lambda_{\rho\alpha},
\end{eqnarray}
has one non-zero component: 
\begin{eqnarray}
R_{00} = \partial_l\Gamma^l_{00}= \partial_l \partial^l \Phi.
\end{eqnarray}
Combining the expression for the Ricci tensor with Gauss' law $\partial_i g^i =- 4 \pi \rho$, we obtain the relationship between matter density and 
the curvature of spacetime in Newton-Cartan theory: 
\begin{eqnarray}
\label{eq:riccipeq}
- \partial_i g^i = R_{00} = 4 \pi \rho.
\end{eqnarray}
Note that there are now two ways to define $\rho$:  through Gauss' law $\partial_i g^i = -4\pi \rho$ and through the curvature equation $R_{00} = 4 \pi \rho$.
In inertial frames these conditions are one and the same. As we will show,  \eref{eq:riccipeq} also holds in linearly accelerating reference frames 
but not in rotating reference frames where Gauss' law defines a different $\rho'$. 

Evidently, \eref{eq:riccipeq} is the Newton-Cartan analogue of Einstein's field equations. In fact, following the discussion from \ref{app:2}, substituting the result for $R_{00}$ into the Einstein field equation \eref{eq:eeq} and using $T_{ij}=T_{0j}=0$, $T_{00}=\rho$, $T=g^{\mu\nu}T_{\mu\nu}$, and $g^{\mu\nu}=\left(g_{\mu\nu}\right)^{-1}$ gives Poisson's equation \eref{eq:ngpeq}. This shows the reduction of Einstein's theory \eref{eq:eeq} to Newton's in the appropriate energy regime, as alluded to at the end of \ref{app:2}.  

However, despite the geometrization of Newtonian gravity and the reduction of GR to that geometric theory, there is a fundamental difference between GR 
and Newton-Cartan theory in that the latter is not a metric theory. As seen below, in contrast to Poincar\'e transformations, Galilean transformations 
cannot be defined as those that leave a metric tensor invariant. Hence, the Newton-Cartan manifold is neither a Riemannian manifold $V_4$, nor its 
generalization (to include torsion) $U_4$. In particular, the connection \eref{eq:ncc} is not given by the Christoffel symbol $\left\{\begin{array}{c}\lambda\\ \mu\nu\end{array}\right\}$ computed 
from a metric. It is for this reason that we have referred to \eref{eq:gdeq} as the autoparallel, rather than geodesic, equation, as it follows directly from the general definition 
of affine connection and parallel transport of a vector $\vec A=A^\mu\partial_\mu=\frac{dx^\mu}{d\tau}\frac{\partial}{\partial x^\mu}$, acting as a differential operator on a smooth manifold. The differential change of the components of this vector is given by $dA^\mu=-\Gamma^\mu_{\alpha\beta}A^\alpha dx^\beta$. Substituting the definition for $\vec A$ into the parallel transport equation yields \eref{eq:gdeq} after a simple exercise in calculus and index gymnastics. In this light, there is no \emph{a priori} reason to demand that the Newton-Cartan connection be symmetric in its lower indices, although it is only the 
symmetric part of the connection that enters the autoparallel equation. Of course, the only non-vanishing component \eref{eq:ncc} is clearly symmetric. 
\section{Galilean transformations of Newton-Cartan Theory}\label{sec:NCGroup}
In this section we  introduce the Galilei group and its generalization, the Galilean line group, and consider the transformation structure  of Newton-Cartan theory under these groups.
Our main conclusion is that in a rotating reference frame, there emerges a gravitomagnetic field that obeys Maxwell's equations and a test particle moves in this field according to the Lorentz force law. 
Furthermore, rather like the construction of \cite{Dyson}, the two inhomogeneous Maxwell equations define a current density $\vec{J}'$ and a matter density $\rho'$, which differs from \eref{eq:riccipeq} by a term that is a coordinate, i.e., gauge, effect. 

The Galilei group $\mathcal{G} = \{ (A, \mathbf{v}, \mathbf{a}, b) \}$, where $A$ is a rotation matrix, $\mathbf{v}$ is a velocity boost, $\mathbf{a}$ is a spatial translation and $b$ is a time translation, 
is a group of spacetime symmetries  under the composition rule
\begin{eqnarray}
&&(A_2, \mathbf{v}_2, \mathbf{a}_2, b_2) (A_1, \mathbf{v}_1, \mathbf{a}_1, b_1) =\nonumber\\ 
&&\ (A_2 A_1, \mathbf{v}_2+A_2\mathbf{v}_1, \mathbf{a}_2 + A_2 \mathbf{a}_1 + b_1\mathbf{v}_2, b_2+b_1).\label{1}
\end{eqnarray}
The inverse of $(A,\mathbf{v},\mathbf{a},b)$ under \eref{1} is given by
\begin{equation}
(A, \mathbf{v}, \mathbf{a}, b)^{-1} = (A^{-1}, -A^{-1}\mathbf{v}, -A^{-1}(\mathbf{a}-b\mathbf{v}), -b).
\end{equation}
The action of a Galilean coordinate transformation $(A, \mathbf{v}, \mathbf{a}, b) \in \mathcal{G}$ on a spactime point $(\mathbf{x},t)$ is defined by
\begin{equation}
\label{GalileanTrans}
(A, \mathbf{v}, \mathbf{a}, b):
\left(\begin{array}{cc}
\mathbf{x}\\
t
\end{array}\right)
\to
\left(\begin{array}{cc}
\mathbf{x}'\\
t'
\end{array}\right)
=
\left(\begin{array}{cc}
A\mathbf{x} + \mathbf{v}t+ \mathbf{a}\\
t + b
\end{array}\right).
\end{equation}

The Galilei group ties together all \emph{inertial} reference frames in a Galilean spacetime. The \emph{Galilean line group} generalizes the Galilei group to include 
transformations into all noninertial reference frames. This can be done by demanding rotations $A$ and space translations $\mathbf{a}$ of \eref{GalileanTrans} be arbitrary functions of time. 
Thus, consider 
\begin{eqnarray}
\label{GalileanLineTrans}
(A, \mathbf{a},b):
\left(\begin{array}{cc}
\mathbf{x}\\
t
\end{array}\right)
\to
\left(\begin{array}{cc}
\mathbf{x}'\\
t'
\end{array}\right)
=
\left(\begin{array}{cc}
A(t)\mathbf{x} + \mathbf{a}(t)\\
t+b
\end{array}\right).
\end{eqnarray}
From this, we deduce the composition rule for the set of transformations $\mathbb{G}:= \{(A,\mathbf{a}, b)\}$: 
\begin{eqnarray}
\label{CompositionRule}
&&(A_2,\mathbf{a}_2,b_2)(A_1,\mathbf{a}_1,b_1) =\nonumber\\
&&\quad((\Lambda_{b_1}A_2)A_1, (\Lambda_{b_1}A_2)\mathbf{a}_1 + \Lambda_{b_1}\mathbf{a}_2, b_1+b_2)
\end{eqnarray}
where $\Lambda$ is the shift operator $\Lambda_b f(t) = f(t+b)$. It accounts for the fact that in successive application two transformations, the 
$A_2$ and $\mathbf{a}_2$ of the second group element are to be evaluated at $t+b_1$, whereas the $A_1$ and $\mathbf{a}_1$ of the first element
are evaluated at $t$.  It is straightforward to verify that \eref{CompositionRule} is associative. Further, each element of $\mathbb{G}$ has an inverse under  \eref{CompositionRule}: 
\begin{eqnarray}
(A, \mathbf{a}, b)^{-1} = (\Lambda_{-b}A^{-1},-\Lambda_b(A^{-1}\mathbf{a}), -b).
\end{eqnarray}
Therefore, $\mathbb{G}$ is a group.
We refer to it as the Galilean line group. A more complete analysis of this group can be found in \cite{2012AnPhy.327.2310M}.

When we set  $A(t)=A$ to be a constant rotation and $\mathbf{a}(t) = \mathbf{v}t+\mathbf{a}^{(0)}$, both the Galilean line group element $(A(t),\mathbf{a}(t),b)$ and the corresponding Galilean group element $(A, \mathbf{v}, \mathbf{a}^{(0)},b)$ have the same action on all spacetime points $(\mathbf{x},t)$ (see \eref{GalileanTrans} and \eref{GalileanLineTrans}).
Thus $\mathcal{G}$ is isomorphic to a subgroup of $\mathbb{G}$, i.e., the Galilean line group generalizes the Galilei group.

Given coordinate transformations \eref{GalileanLineTrans}, we can readily compute the transformation rules for the differential operators, therewith arbitrary vector fields, under $\mathbb{G}$:
\begin{eqnarray}
\frac{\partial}{\partial t}&=\frac{\partial {t^\prime}}{\partial t}\frac{\partial}{\partial {t^\prime}}+\frac{\partial{x^\prime}^l}{\partial t}\frac{\partial}{\partial{x^\prime}^l}\nonumber\\
&=\frac{\partial}{\partial t^\prime}+\dot{A}^l_{\ k}x^k\frac{\partial}{\partial{x^\prime}^l}+\dot{a}^l\frac{\partial}{\partial{x^\prime}^l}\label{DiffTrans1}\\
\frac{\partial}{\partial x^k}&=\frac{\partial t^\prime}{\partial{x}^k}\frac{\partial}{\partial t^\prime}+\frac{\partial{x^\prime}^l}{\partial x^k}\frac{\partial}{\partial{x^\prime}^l}
=A^l_{\ k}\frac{\partial}{\partial{x^\prime}^l}\label{DiffTrans2}
\end{eqnarray}
where we have made use of the fact that $A$ and $\mathbf{a}$ are functions of $t$ only, not $\mathbf{x}$. Though the arguments are functions of time, note that the matrices 
$A$ of $\mathbb{G}$ are orthogonal,  $AA^T=I$, a property that we will repeatedly use throughout this paper. By virtue of this orthogonality, 
\eref{DiffTrans2} implies $\nabla^2={\nabla^\prime}^2$. 

Let us now consider the transformation properties of the Newton-Cartan theory under $\mathbb{G}$. All of the relevant information is encoded in the transformation structure of the 
autoparallel equation:
\begin{equation}\label{ModifiedNewton}
\frac{d^2 x^j(\mathbf{x}',t')}{d\tau^2} + \delta^{ij} \frac{\partial x'^k}{\partial x^i}\frac{\partial \Phi(x(\mathbf{x}',t'))}{\partial x'^k} \left(\frac{dt}{d\tau}\right)^2 = 0,
\end{equation}
where $\mathbf{x}'$ and $t'$ are defined by  \eref{GalileanLineTrans}.
It follows  that $\frac{dt'}{dx^i}=0$, which we have made use of in writing \eref{ModifiedNewton}.

In order to calculate $\frac{d^2 x^k}{d \tau^2}$ explicitly, we need the inverse of \eref{GalileanLineTrans}: $x^k = A_j^{\ k}(x'^j - a^j)$.
Expanding the derivative $\frac{d^2x^k}{d\tau^2}$, inserting it in \eref{ModifiedNewton}, and rearranging terms we find 
 \begin{eqnarray}
0 &=& \frac{d^2 x'^l}{d \tau^2} +( 2 A^{lk} \dot{A}_{jk}) \left( \frac{d x'^j}{d\tau}\right) \left( \frac{dt}{d\tau} \right)\nonumber\\
&&+\left( \delta^{lj} \frac{\partial \Phi}{\partial x'^j}- A^{lk}\ddot{A}_{jk} x^j - A^{lk} \frac{d^2}{dt^2} \left[ A_{jk}a^j \right] \right) \left( \frac{dt}{d\tau}\right)^2.\nonumber\\
\end{eqnarray}
Comparing this with the autoparallel equation shows that the only non-zero connection components  are
\begin{eqnarray}
\Gamma'^l_{00} &= \delta^{kl}\frac{\partial \Phi}{\partial x'^k} + A^{lk} \ddot{A}_{jk} x'^j - A^{lk} \frac{d^2}{dt^2} (A_{jk}a^j)\label{GammaPrimes1}\\
\Gamma'^l_{0j} &= \Gamma'^l_{j0} = A^{lk}\dot{A}_{jk}\label{GammaPrimes2}
\end{eqnarray}
Taking time derivative of $A^{lk}A_{jk} = \delta^l_{\ j}$ shows that the connection components \eref{GammaPrimes2} 
are antisymmetric in $l$ and $j$, i.e., $\Gamma'^l_{0j}=\Gamma'^l_{j0} = -\Gamma'^j_{0l} = -\Gamma'^j_{l0}$. 
For an alternative more geometric approach that confirms these results, see \ref{app:1}. 

\subsection{Simulated gravitomagnetic potentials and fields} 
The natural reading is that $\Gamma'^l_{00} = -g'^l$, the gravitational field in the transformed frame. However, it has 
non-zero curl because of the term $A^{lk} \ddot{A}_{jk} x'^j$ so we can no longer write $\vec{g}'$ 
as simply the gradient of a scalar potential. Instead, we can write it as a combination of the gradient of a scalar potential $\Phi'$ and the time derivative of a 
vector potential $\vec{W}'$: 
\begin{equation}
g'^i = -\delta^{ij}\partial'_j \Phi' - \partial'_t W'^i,\label{gprime}
\end{equation}
where 
\begin{eqnarray}
\Phi' &= \Phi - x'^l A_l^{\ k} \frac{d^2}{dt^2}(A_{jk}a^j)\label{Phi'}\\
W'^l &= \int dt' A^{lk}\ddot{A}_{jk}x'^j.\label{W'}
\end{eqnarray}
The introduction of the vector potential readily gives rise to a gravitational magnetic field $\vec{h}'$: 
\begin{equation}
h'^i = \varepsilon^i_{\ jk} \partial'^j W'^k.\label{hprime}
\end{equation}

All of this is in complete parallel with electromagnetic theory, where electrostatic phenomena are described 
by a scalar potential but electrodynamic phenomena require the introduction of a vector potential. The time dependence of the 
transformation matrices in \eref{GalileanLineTrans} makes the theory dynamic, leading to time dependent scalar and vector potentials which 
can be explicitly determined as in \eref{Phi'} and \eref{W'} by the element of $\mathbb{G}$  that implements the transformation to the 
noninertial primed frame. Further, just as in electrodynamics,  $g'^i$ and $h'^i$ exhibit a gravitomagnetic $U(1)$-symmetry,
\begin{eqnarray}
\label{eq:u1sym}
\Phi'\to\Phi'-\partial_{t'}\gamma\\
W'^i\to W'^i+\delta^{ij}\partial_{j}\gamma,
\end{eqnarray}
where $\gamma$ is an arbitrary function.

Note that \eref{gprime} gives us the connection components ${\Gamma'}^i_{00}=-g^i$ in terms of the potentials. 
Likewise, we can express the remaining components  $\Gamma'^i_{0j}$ of the connection also in terms of the potentials:   
 \begin{eqnarray}
\Gamma'^i_{0j}&=&A^{ik}\dot{A}_{jk}\nonumber\\
&=&\frac{1}{2} \int dt' \left\{ A^{ik}\ddot{A}_{jk} - A_{jk}\ddot{A}^{ik} \right\}\nonumber \\
&=&  \frac{1}{2} \left( \partial'_j W'^i -\partial'^i W'_j\right)\nonumber\\
&=&-\frac{1}{2} \varepsilon^i_{\ jk} \varepsilon^k_{\ lm} \partial'^l W'^m
\end{eqnarray}
where we have used the identity $A_{jk}\ddot{A}_i^{\ k} = -2\dot{A}_{ik}\dot{A}_j^{\ k} - A_{ik}\ddot{A}_j^{\ k}$, which follows from $A_{ik}A_j^{\ k} = \delta_{ij}$ upon differentiation, 
and the definition \eref{W'}. 

Further,  from the autoparallel equation,
\begin{equation}
0 = \frac{d^2 x'^i}{d\tau^2} +\Gamma'^{i}_{00} \left( \frac{dt'}{d\tau} \right)^2 + 2 \Gamma'^i_{j0} \left( \frac{d x'^j}{d\tau} \right) \left( \frac{dt'}{d\tau} \right),\nonumber 
\end{equation}
we can extract the force law in the noninertial frame:
\begin{eqnarray}
\frac{d^2{x'}^i}{d{t'}^2} &=& -\Gamma'^i_{00} -2\Gamma'^{i}_{j0} v'^j\nonumber\\
&= &g'^i + \varepsilon^i_{\ jk} v'^j h'^k\label{LorentzForce}
\end{eqnarray}
where $v'^i$ is the velocity of the test particle. 

We can also express \eref{LorentzForce} in terms of angular velocity $\boldsymbol{\omega}$ 
to obtain the familiar expression for the acceleration of a particle in a rotating reference frame: 
\begin{equation}
\frac{d^2{\mathbf{x}'}}{d{t'}^2}=-\nabla\Phi'-\boldsymbol{\omega}\times(\boldsymbol{\omega}\times \mathbf{x}')-\dot{\boldsymbol{\omega}}\times\mathbf{x}'-2\boldsymbol{\omega}\times\mathbf{v}'.\label{omega}
\end{equation}
Here, we have used the familiar identity $A^i_{\ k}\dot{A}_j^{\ k}{x'}^j=\varepsilon^i_{\ jk}\omega^j{x'}^k$ or, its component-free form
$A\dot{A}^T\mathbf{x}'=\boldsymbol{\omega}\times\mathbf{x}'$, and its derivative which gives $A\ddot{A}^T\mathbf{x}'=
\boldsymbol{\omega}\times(\boldsymbol{\omega}\times \mathbf{x}')+\dot{\boldsymbol{\omega}}\times\mathbf{x}'$. 
Hence, we see that the second, magnetic term of \eref{LorentzForce} is simply the Coriolis force while 
the second term of \eref{GammaPrimes1} is the sum of centrifugal and Euler terms, which are encapsulated in the gravitational part $\boldsymbol{g}'$ 
of \eref{LorentzForce}. The term $A\frac{d^2}{dt^2}\left(A^T\mathbf{a}\right)$, which describes the effects of a linear 
acceleration of the reference frame, is absorbed into the scalar potential $\Phi'$. 
It is a remarkable feature of Cartan's geometric formulation that these different noninertial effects are naturally encoded in different components of the connection, 
i.e., those that depend on the position in the ${\Gamma'}^i_{00}$-components and those that depend on velocity in the ${\Gamma'}^i_{0j}$-components. This splitting of noninertial 
effects brings to light a very interesting property of \eref{LorentzForce} that we  do not easily see in the more commonly known form \eref{omega}: the equation of motion of a nonrelativistic particle in a noninertial reference frame has the same form as the equation of motion under the relativistic Lorentz force!

Not surprisingly,  the simulated gravitational and magnetic fields are governed by Maxwell's equations.
Equations \eref{gprime} and \eref{hprime}, which relate the fields to their potentials, automatically guarantee the two homogeneous Maxwell equations:
\begin{eqnarray}
\partial_i h'^i &= 0\\
\varepsilon^i_{\ jk} \partial'^j g'^k &= -\partial'_t h'^i.
\end{eqnarray}
The inhomogeneous Ampere's law and Gauss' law, which can be considered the definitions of $\rho'$ and $J'^i$, read
\begin{eqnarray}
\partial'_i g'^i &= -4\pi \rho'\label{Gauss}\\
\varepsilon^i_{\ jk} \partial'^j h'^k &= -4\pi J'^i + \partial'_t g'^i\label{Ampere}
\end{eqnarray}
where the signs of $\rho'$ and $J'$ have been reversed from electrictrodynamics because in gravity like charges attract instead of repel.
Note that Gauss' and Ampere's laws automatically guarantee the continuity equation,
\begin{equation}
\partial'_i J'^i +\partial'_t \rho' = 0.
\end{equation}
This, in conjunction with \eref{eq:u1sym}, is the low energy equivalent of energy, momentum, and stress conservation generated by the full diffeomorphism symmetry in GR.

From Guass' and Ampere's laws we obtain explicit expressions for matter and current densities: 
\begin{eqnarray}
\rho'_{Gauss} &=& \frac{1}{4\pi} \partial'_l \left(-g'^l \right) =\frac{1}{4\pi} \partial'_l \left(\Gamma'^l_{00} \right)\nonumber\\
&= &\rho + \frac{1}{4\pi} A^{lk}\ddot{A}_{lk}= \rho - \frac{1}{4\pi} \dot{A}^{lk}\dot{A}_{lk}\label{rhoGauss}\\
J'^i &= &-\frac{1}{4\pi} \left( \varepsilon^i_{\ jk} \varepsilon^k_{\ lm}\partial'^j \partial'^l W'^m + \partial'_t \Gamma'^i_{00} \right)\nonumber\\
&= &-\frac{1}{4\pi} \partial'_t \Gamma'^i_{00}.\label{Jprime}
\end{eqnarray}
where in the expression for $\rho'_{Gauss}$ we made use of the fact that two time derivatives of $A^{lk}A_{lk}=4$ gives $A^{lk}\ddot{A}_{lk} = - \dot{A}^{lk} \dot{A}_{lk}$.
The subscript `Gauss' in  \eref{rhoGauss} is to indicate explicitly that this $\rho'$ is defined by way of Gauss' law. Going back to \eref{eq:riccipeq}, 
we see that we can also define a matter density in terms of the Ricci curvature equation, a possibility to which we now turn. As will be shown below, the equality  
\eref{eq:riccipeq} fails in a rotating reference frame, leading to a schism between geometry and Gauss' law. 

\subsection{Geometry and the Ricci field equation}
A direct computation shows that, as before, only the $R'_{00}$ component  is non-zero in noninertial frames and that it is in fact invariant: 
\begin{eqnarray}
\label{RicciPrime}
R'_{00} &= \partial'_l\Gamma'^l_{00} - \partial'_t \Gamma'^l_{l0} - \Gamma'^l_{0k}\Gamma'^k_{l0} \nonumber\\
&= \partial'_l\Gamma'^l_{00} + \Gamma'^l_{k0}\Gamma'^l_{j0} \delta^j_{\ k} \nonumber\\
&= 4\pi \rho - \dot{A}^{ki} \dot{A}_{ki} + A_{lj}\dot{A}^{kj} A^{li}\dot{A}_{ki}\\
&= 4\pi \rho =  R_{00},\nonumber
\end{eqnarray}
where we have used \eref{rhoGauss} and the antisymmetry of $\Gamma'^l_{j0}$ in $l$ and $j$. 
The invariance of $R_{00}$ shows that the Ricci tensor is in fact a tensor under $\mathbb{G}$, 
providing support for our claim  in the introduction that the geometric content of Newton-Cartan gravity is invariant 
under $\mathbb{G}$. In particular, that the curvature of a Newtonian spacetime is  determined 
by the matter density holds true both in inertial and noninertial reference frames. 

\subsection{Gauss vs Ricci}

It readily follows from \eref{rhoGauss} and \eref{RicciPrime} that 
\eref{eq:riccipeq} is not generally satisfied in noninertial frames, leading to 
\begin{equation}
\rho = \rho'_{Ricci} \neq \rho'_{Gauss}.
\end{equation}
In order to appreciate this discrepancy between Gauss' law and the Ricci field equation, we must observe that the $\rho$ that appears in the latter is the real, physical matter density of the world. 
In contrast, since $\dot{A}^{lk}\dot{A}_{lk}$ is spatially constant, the mass $M'_{Gauss} = \int_{\mathbb{R}^3}d^3x\,\rho'_{Gauss}$ associated with the Gaussian density diverges.
Similarly, the current density $J'^i$ contains a term $\left( \frac{1}{4\pi} \partial'_t [A^{ik} \ddot{A}_{jk}x'^j]\right)$ that increases with distance from the origin and 
a term $\left( \frac{1}{4\pi}\partial'_t [A^{ik} \partial_t^2(A_{jk}a^j)] \right)$ that remains spatially constant so the total current entering 
space $\oint_{\partial \mathbb{R}^3} d\mathbf{A} \cdot \mathbf{J'}$ diverges as well. In other words, there exist no physical matter and charge densities that can generate 
gravitomagnetic fields exactly equivalent to those  that appear in rotating reference frames. It is for this reason that we have referred to them as simulated fields. 

Even though the densities $\rho'_{Gauss}$ and $\rho'_{Ricci}$ are not equal and  $\rho'_{Gauss}$ does not correspond to a finite mass, it is important to recognize that they both lead to the same autoparallel equation of motion for a test particle. Thus, to the extent that what is experimentally accessible are only the trajectories of test particles, the schism between the geometric Ricci equation and dynamical Gauss equation has no experimental consequences. In this sense, we can consider the discrepancy between $\rho=\rho^\prime_{Ricci}$ and $\rho^\prime_{Gauss}$ as a mass gauge. In fact, it is possible to generate an effective mass density $\rho'_{Gauss}$ purely through rotations with no physical mass by, for instance, stepping into a rotating frame such as a merry-go-round.

The above discussion also illustrates the striking difference between linear accelerations and rotational accelerations. Note that $\rho'_{Gauss}$  differs from the original matter density $\rho$ by terms that  depend only on time dependent rotations and their derivatives; linear accelerations never lead to a redefinition of 
$\rho$ and the Gauss law and Ricci field equation live harmoniously in linearly accelerating reference frames. Explicitly, 
if only linear accelerations are present, i.e., $\dot{A}_{ij}=0$, then
\begin{eqnarray}
\label{linearacc}
\cases{
{\Gamma'}^l_{00}&$=\delta^{lk}\frac{\partial\Phi'}{\partial{x'}^k}-\ddot{a}^l$\\
{\Gamma'}^l_{0j}&$=0$\\
\Phi'&$=\Phi-\mathbf{x}'\cdot\ddot{\mathbf{a}}$\\
W'^i &$= 0$\\
\rho_{Gauss}'&$=\rho=\rho'_{Ricci}$
}
\end{eqnarray}
These equations tell us that under linear accelerations the autoparallel equation remains form invariant and the potential picks up an additional term that is linear in position so that the Poisson equation remains form invariant. If we further restrict ourselves to ordinary Galilean transformations, then $\ddot{\mathbf{a}}=0$ so that ${\Gamma'}^l_{00}=\Gamma^l_{00}$ and 
$\Phi'=\Phi$. In a very real sense, the difference between linearly and rotationally accelerating reference frames is far greater than that between linearly accelerating and inertial reference frames. 

As a final remark, given that both $\rho'_{Gauss}$ and $\rho'_{Ricci}$ lead to the same autoparallel equation, it is natural to ask if we can bridge the divide between geometry and Gauss' law, i.e., restore $R'_{00} = -\partial'_i g'^i$, by defining a covariant derivative to replace $\partial'_i$.
However, a brief calculation using the Newton-Cartan connection $\Gamma$  shows that such a covariant derivative does not resolve the situation. 
\section{Effective Gravitating Mass and ADM-Chrage}\label{sec4}
In the foregoing discussion, we have shown that time dependent rotations give rise to an effective gravitating mass density $\rho'_{Gauss}$, which is different from the Ricci density $\rho'_{Ricci}$ that 
remains invariant and equal to the physical matter density $\rho$. Given this discrepancy and the emergence of effective gravitating mass distributions, it remains relevant to discuss how to define (a proper) mass within a Newton-Cartan formulation that is universal irrespective of the frame of reference.  We also explore how this mass relates to the ADM (R.~Arnowitt, S.~Deser and C.W.~Misner) mass of general relativity. To accomplish this, given that the simulated magnetic field that emerges in rotating reference frames is at the heart of the discrepancy between $\rho'_{Gauss}$ and $\rho'_{Ricci}=\rho$, we consider the gravitational field effects and effective mass that arises from a matter density  endowed with a pure electric charge. It is not obvious that a simple electric charge should alter Newton's universal gravitational force law \eref{eq:fng}. However, as seen below, the remnants of the contribution of the electric charge to the stress-energy tensor $T_{\mu\nu}$ of GR do lead to an effective mass. Nonetheless, there are still important differences between this effective mass and  $\rho'_{Gauss}$ discussed above, further highlighting the difficulty of mirroring rotational effects by physical mass distributions. 

To begin, we will make use of the Reissner-Nordstr\"om solution of GR and consider a line element with static potential, as in App.~\ref{app:2}:
\begin{eqnarray}
\label{eq:rns}
ds_{RNS}^2=-f(r)dt^2+\frac{dr^2}{f(r)}+r^2d\Omega^2,
\end{eqnarray}
where now $f(r)=1-\frac{2M}{r}+\frac{k Q^2}{r^2}$ and $k=\frac{1}{4\pi \epsilon_0}$. This line element describes a gravitating charged point mass $\rho=M\delta^3(r)$ with Coulomb potential:
\begin{eqnarray}
\label{}
\Phi_C=-\frac{Q}{4\pi \epsilon_0 r}.
\end{eqnarray}
In contrast to the Schwarzschild solution \eref{eq:sss}, the Reissner-Nordstr\"om metric is not a vacuum solution and solves the Einstein field equation:
\begin{eqnarray}
\label{}
R_{\mu\nu}=8\pi T^{EM}_{\mu\nu},
\end{eqnarray}
where $T^{EM}_{\mu\nu}$ is the covariant electromagnetic energy momentum tensor. To formulate a covariant electromagnetic theory, the standard procedure is to define the 
four vector potential $A_\mu=\left\{-\Phi_C,\vec A\right\}$, which gives rise to the electromagnetic field strength tensor:
\begin{eqnarray}
\label{}
F_{\mu\nu}=\partial_\mu A_\nu-\partial_\nu A_\mu.
\end{eqnarray}
It is easy to see from the definition of $A_\mu$ that $F_{\mu\nu}$ encodes the electric and magnetic field in the following way:
\begin{eqnarray}
\label{}
F_{0i}=-E_i~{and}~F_{ij}=\epsilon\indices{_{ij}^k}B_k
\end{eqnarray}
and is invariant with respect to $U(1)$-gauge transformations of the form $A_\mu\to A_\mu+\partial_\mu\Lambda$. In this formulation the electromagnetic field energy momentum tensor takes the form:
\begin{eqnarray}
\label{eq:ememt}
T^{EM}_{\mu\nu}=\frac{1}{4\pi k}\left(F_{\alpha\mu}F\indices{^\alpha_\nu}-\frac14g_{\mu\nu}F_{\alpha\beta}F\indices{^\alpha^\beta}\right).
\end{eqnarray}
Clearly $T^{EM}=g^{\mu\nu}T^{EM}_{\mu\nu}=0$ since by definition of the inverse metric, $g^{\mu\nu}g_{\mu\nu}$ must be equal to the respective dimension of spacetime. The vanishing of $T^{EM}$ is therefor only true in four dimensions and signals an additional property of electromagnetic theory, in that it is conformally invariant in four spacetime dimensions. 

Next, using \eref{eq:neem}, we obtain our Newtonian gravitational field and field strength for a charged point mass:
\begin{eqnarray}
\label{}
\Phi=&-\frac{M}{r}+\frac{kQ^2}{2 r^2}\\
\vec g=&-\frac{M}{r^2}+\frac{kQ^2}{ r^3}.
\end{eqnarray}
We should note that endowing the mass density with an electric charge still yields a static theory and hence there are no gravitomagnetic contributions. However, the added charge $Q$ alters the Gauss law constraint to include an effective mass contributing to the field strength flux through some Gaussian-Sphere:
\begin{equation}
\label{}
\oint_{GS}\vec g\cdot d\mathbf{A}=\left(-\frac{M}{r^2}+\frac{kQ^2}{r^3}\right)4\pi r^2=-4\pi M_{eff}
\end{equation}
where the effective gravitating mass $M_{eff}$ is
\begin{eqnarray}
\label{eq:effmem}
M_{eff}=M-\frac{kQ^2}{r}.
\end{eqnarray}
Equivalently, we could evaluate $\nabla^2 \Phi$ to obtain $\rho_{eff}=M\delta^3(r)-\frac{kQ^2}{4\pi r^4}$ and then integrate over all of space to yield the enclosed mass. However this approach is a bit cumbersome as we encounter an ultraviolet (small $r$) divergence due to the $1/r^4$ behavior in the effective mass density. Introducing a proper uv-cutoff reproduces \eref{eq:effmem} and signals that our chosen approach to gravity is incomplete. We should also note that this same uv-divergence is still present in GR and shows up as a proper curvature singularity of \eref{eq:rns}. The fact that we did not encounter the uv-divergence in the Gauss law constraint is due to the fact that this constraint measures asymptotic behavior of the field strength. Additionally, we know from singularity theorems and the laws of black hole mechanics that the only mass parameter of \eref{eq:rns} at asymptotic infinity should be completely given by its ADM-mass\cite{MTW}, which in the Reissner-Nordstr\"om case reads $M_{ADM}=M$. This motivates us to define an analogous parameter, which we will call the Newtonian-ADM mass, by:
\begin{eqnarray}
\label{eq:nadm}
M_{NADM}=\frac{1}{4\pi}\lim_{r\to\infty}\oint\vec g\cdot d\mathbf{A}.
\end{eqnarray}
It is clear from \eref{eq:effmem} that $M_{ADM}=M_{NADM}$ and implies that the gravitational mass in Newton's gravity is the same as the $ADM$ mass of GR. We can make this equivalence concrete by considering spacetimes that exhibit asymptotically flat time symmetric initial data, i.e., $g_{ij}\approx\delta_{ij}+\mathcal{O}\left(\frac{1}{r}\right)$. For such spacetimes, the $ADM$ mass reads \cite{MTW,Brewin:2006qe}:
\begin{eqnarray}
\label{eq:admmass}
M_{ADM}=\frac{1}{16\pi}\lim_{r\to\infty}\oint\delta^{ij}\left(\partial_ig_{jk}-\partial_kg_{ij}\right)n^k dS.
\end{eqnarray}
Here, $dS$ is a topological two sphere with unit normal $n^k$. Obviously,  \eref{eq:admmass} is not a covariant statement, but it is evaluated for asymptotic Euclidian coordinates. For spacetimes of the form \eref{eq:sss} and \eref{eq:rns}, we can obtain the asymptotically flat time symmetric initial data metric by setting $dt=0$, Taylor expanding, and performing radial redefinitions to obtain in both cases:
\begin{eqnarray}
\label{eq:gij}
ds^2_{3}=g_{ij}dx^idx^j=g(r)\left(dx^2+dy^2+dz^2\right),
\end{eqnarray}
where $g(r)=1-2\Phi+\mathcal{O}\left(\frac{1}{r^2}\right)$. Using this in \eref{eq:admmass}, we obtain:
\begin{eqnarray}
\label{}
&\delta^{ij}\left(\partial_ig_{jk}-\partial_kg_{ij}\right)n^k=-4\partial_i\Phi n^i\Rightarrow\nonumber\\
M_{ADM}&=\frac{1}{16\pi}\lim_{r\to\infty}\oint\delta^{ij}\left(\partial_ig_{jk}-\partial_kg_{ij}\right)n^k dS\nonumber\\
&=\frac{1}{16\pi}\lim_{r\to\infty}\oint-4\partial_i\Phi n^i dS\nonumber\\
&=\frac{1}{4\pi}\lim_{r\to\infty}\oint\vec g\cdot d\mathbf{S},
\end{eqnarray}
validating our formula \eref{eq:nadm} for the appropriate choice of a Gaussian surface. 

If we turn our attention back to \eref{eq:effmem}, we see that there exists a point 
\begin{eqnarray}
\label{eq:0g}
r_{zero}=\frac{kQ^2}{M},
\end{eqnarray}
where the effective mass vanishes, i.e., a point of zero gravity. While this is an interesting result, since $r_{zero}$ depends on the ratio between charge squared and mass, it will be physically constrained by the coordinate singularity (horizon) where $f(r_{\pm})=0$ of \eref{eq:rns}:
\begin{eqnarray}
\label{eq:rpm}
r_{\pm}=M\pm\sqrt{M^2-kQ^2}.
\end{eqnarray}
From the above we see that the amount of charge that $M$ may carry is bounded above (extremal limit) by the radical term in $r_{\pm}$ to be:
\begin{eqnarray}
\label{eq:maxQ}
\frac{M}{\sqrt{k}}>Q.
\end{eqnarray}
This constraint pushes any physical value of $r_{zero}$ close to the uv-divergent regime of our theory and past the point of the coordinate singularity located at $r_{\pm}$. Not to say that this regime of zero gravity, brought about by charging up $M$, does not exist, but it warrants a more in-depth analysis of the effective mass within a uv-complete theory of gravity. 

Another interesting feature of the charged mass distribution comes from its Newton-Cartan reformulation. Comparing Section~\ref{sec:NCGroup} to \ref{app:2} for the Schwrazschild spacetime, we find an exact agreement between Einstein and Newton. This is not the case for Reissner-Nordstr\"om, in fact we only find agreement up to $\mathcal{O}\left(1/r^5\right)$ in $R_{\mu\nu}$, which  is clearly insignificant for large radial distances. This discrepancy stems from the fact that we are not dealing with a vacuum ($T_{\mu\nu}=0$ except at $r=0$) solution and the higher order contributions in $R_{\mu\nu}$ originate from the electromagnetic field in \eref{eq:ememt}. Looking back at \eref{GammaPrimes1} we see that due to the time dependent rotations our definition for the NADM mass, \eref{eq:nadm}, is now plagued by infrared divergent terms. Scenarios of this type are known from non-asymptotically flat  solutions in GR and require a redefinition of the ADM mass in terms of $T_{00}$ and techniques from holographic regularization to cancel infrared divergences. In this procedure, boundary counter terms are added to the bulk Lagrangian and invoking the Hamilton-Jacobi variational principle yields a finite $ADM$ mass\cite{Liu:2004it}. However, we are not sure how to implement a similar program in the present case, since counter terms in general are motivated from the action principle of GR, which is not applicable in a Newton-Cartan formulation. But, we can implement an analogous definition of the ADM mass in terms of $T_{00}$, since from \eref{RicciPrime} we have:
\begin{eqnarray}
\label{eq:newnadm}
\lim_{r\to\infty}\frac{1}{4\pi}\int R_{00}dV=\frac{1}{4\pi}\lim_{r\to\infty}\oint\vec g\cdot d\mathbf{A}=M_{NADM}.
\end{eqnarray}
The left side of the above equation provides us with a more robust and purely geometric definition of ADM mass since $R'_{00}=R_{00}$ and it alleviates the need for any mass regularization. This definition only applies within a Newton-Cartan formulation and implies that any effective matter distributions should exhibit conformal invariance.  Also, in contrast to \eref{eq:nadm}, it leaves us with a covariant way of computing the ADM mass. 
\section{Concluding Remarks}\label{sec5}
In this paper, we have studied the structure of Newton-Cartan gravity under coordinate transformations into both linearly and rotationally accelerating reference frames. We have shown that  Cartan's interpretation of Newtonian gravity as a geometric theory is tenable in accelerating reference frames in the sense that the Ricci field equation that connects the curvature of spacetime with the matter density remains an invariant scalar equation under the action of the entire Galilean line group. However, there are some very interesting differences between the subgroups of linear acceleration transformations and rotational acceleration transformations. While the gravitational field (as well as matter density) becomes time dependent under either linear or rotational accelerations, the structure of the theory in linearly accelerating reference frames is more or less the same as that in inertial reference frames. In contrast, when transformed to a rotating reference frame, there appears a simulated magnetic field of gravitational origin, leading to a set of field equations of the same form as Maxwell's equations in electrodynamics.  The simulated gravitomagnetic field produces a Coriolis force on test particles while the centrifugal and Euler terms in the connection generate a simulated gravitational field.  In rotationally accelerating frames, the matter and current densities defined by way of the gravitomagnetic Maxwell equations do fulfill the continuity equation, but they are unphysical in the sense that their spatial integrals diverge and do not define a finite mass or current. Nevertheless, the equation of motion for a test particle is consistent with the Ricci field equation and the physical matter density.  We have addressed the unphysical nature of the matter density by showing how to compute the ADM mass in a covariant and robust manner within this formulation. 

As noted in the introduction, perhaps the most remarkable feature of this study is that we start with Newton's gravity, a Galilean covariant theory, and by performing a set of coordinate transformations which are themselves generalizations of Galilean transformations, arrive at a theory that has the same form as electrodynamics, well-known to be covariant with respect to special relativity and $U(1)$ gauge symmetry. In this regard, it bears a striking parallel to Dyson's study of Feynman's proof of Maxwell's equations from Newton's law of motion and the Heisenberg commutation relations \cite{Dyson}.
\ack{L.R. is grateful to Grinnell College for three years of support and mentorship, and to NASA Goddard Space Flight Center for their hospitality. This work was supported in part by the HHMI Undergraduate Science Education Award 52006298 and the Grinnell College Academic Affairs' CSFS and MAP programs.
}
\appendix
\section{Geometrically Motivated Transformation of $\Gamma$}\label{app:1}
In Section~\ref{sec:NCGroup}, we transformed the autoparallel equation and read off $\Gamma'$ in the primed frame. 
Another way to obtain $\Gamma'$ is through a direct transformation, treating it as an affine connection. The purpose of this appendix 
is to show that these two methods of transforming $\Gamma$ give exactly the same result.

The direct transformation law of the connection is obtained by requiring that the covariant derivative of a vector
 $\nabla_\mu V^\nu = \partial_\mu V^\nu + \Gamma^{\nu}_{\mu\lambda} V^\lambda$ transforms as a tensor. 
 This leads to the following transformation condition on the connection (see \cite{Carroll} equation 3.10)
\begin{eqnarray}\label{DirectGamma}
\Gamma'^\nu_{\mu\lambda} =& \frac{\partial x^\alpha}{\partial x'^\mu} \frac{\partial x^\gamma}{\partial x'^\lambda} \left( \frac{\partial x'^\nu}{\partial x^\beta} \Gamma^\beta_{\alpha \gamma} - \frac{\partial^2 x'^\nu}{\partial x^\alpha \partial x^\gamma} \right).
\end{eqnarray}
We will now show that transforming the autoparallel equation leads to the same condition on $\Gamma'$. Recall the autoparellel equation and its transformed counterpart:
\begin{eqnarray}\label{UnprimedGeoEq}
0 =& \frac{d^2 x^\mu}{d \tau^2} + \Gamma^\mu_{\rho\sigma} \frac{dx^\rho}{d\tau} \frac{dx^\sigma}{d\tau}\\
0 =& \frac{d^2 x'^\mu}{d \tau^2} + \Gamma'^\mu_{\rho\sigma} \frac{dx'^\rho}{d\tau} \frac{dx'^\sigma}{d\tau}\label{PrimedGeoEq}.
\end{eqnarray}
Expanding derivatives gives
\begin{eqnarray}
\frac{d^2 x'^\mu}{d\tau^2} =  \frac{d^2x^\nu}{d\tau^2} \frac{\partial x'^\mu}{\partial x^\nu}+ \frac{dx^\alpha}{d\tau}  \frac{dx^\beta}{d\tau}  \frac{\partial^2 x'^\mu}{\partial x^\alpha \partial x^\beta}.
\end{eqnarray}
Using this in \eref{PrimedGeoEq} and multiplying by $\partial x^\lambda/\partial x'^\mu$ gives
\begin{eqnarray}
0 =& \frac{d^2x^\lambda}{d\tau^2}\nonumber\\
&+ \frac{\partial x^\lambda}{\partial x'^\mu}\left(\frac{\partial^2 x'^\mu}{\partial x^\alpha \partial x^\beta} + \Gamma'^\mu_{\rho \sigma} \frac{\partial x'^\rho}{\partial x^\alpha} \frac{\partial x'^\sigma}{\partial x^\beta} \right) \frac{dx^\alpha}{d\tau} \frac{dx^\beta}{d\tau}.
\end{eqnarray}
Comparing this to the unprimed autoparallel equation \eref{UnprimedGeoEq} and rearranging gives \eref{DirectGamma}.
Thus directly transforming $\Gamma$ produces the same result as transforming the autoparellel equation.
\section{From Einstein to Newton}\label{app:2}
In the creation of the most widely accepted and successful theory of gravity, general relativity, Einstein paid close attention to encode Newton's gravity as a limiting theory at low energy scales. To see this embedding, consider the Schwarzschild line element:
\begin{eqnarray}
\label{eq:sss}
ds^2=g_{\mu\nu}dx^\mu dx^\nu=-f(r)dt^2+\frac{dr^2}{f(r)}+r^2d\Omega^2,
\end{eqnarray}
where $f(r)=1-\frac{2M}{r}$, $d\Omega^2=d\theta^2+\sin^2\theta d\varphi$ is the unit sphere line element. The metric that follows from \eref{eq:sss} is a vacuum solution to the Einstein field (Euler-Lagrange) equation of GR
\begin{eqnarray}
\label{eq:eeq}
R_{\mu\nu}=8\pi\left(T_{\mu\nu}-\frac12 g_{\mu\nu}T\right),
\end{eqnarray}
i.e., the Ricci curvature tensor is flat ($R_{\mu\nu}=0$) everywhere except at the origin, where the Schwarzschild solution exhibits a curvature singularity. Physically this implies a point mass $M$ at rest located at the origin and zero matter ($T_{\mu\nu}=0$) elsewhere. Now, knowing that the Newtonian potential $\Phi$ for a point mass $M$ is given by $\Phi=-\frac{M}{r}$, 
we see that Newton's theory is contained in Einstein's  be way of
\begin{eqnarray}
\label{eq:neem}
f(r)=1+2\Phi,
\end{eqnarray}
or, more generally, for spherically symmetric and asymptotically flat spacetimes:
\begin{eqnarray}
\label{eq:newtp}
\Phi=\frac{1}{2}\left(g_{00}-1\right).
\end{eqnarray}
This identification ensures that the Einstein field equations of GR reduce to the Poisson equation of Newton's gravity in the point particle case at low energies. 
The boundary conditions implemented to solve the Einstein field equations yielding the solution \eref{eq:sss} are encoded in Newton's gravity for the point particle case. 
To see this let us evaluate the Laplace equation for the point particle of mass $M$:
\begin{eqnarray}
\nabla^2\Phi=-M\nabla^2\left(\frac{1}{r}\right) =4\pi M\delta^3(r),
\end{eqnarray}
where we have used the fact that $1/r$ is proportional to the Green's function of $\nabla^2$. This implies a mass density $\rho=M\delta^3(r)$, which is precisely the condition set forth on $T_{\mu\nu}$ for the Schwarzschild solution, in that $T_{ij}=T_{0j}=0$ and $T_{00}\sim\rho$ at the origin and zero everywhere else. This implies an equivalence between the the Poisson equation for Newton's gravity and the zero-zero component of the Einstein field equation.
\vspace{.5cm}
\begin{center}
\noindent\line(1,0){150}
\end{center}
\bibliographystyle{unsrt}
\bibliography{GGNC}

\end{document}